%% file: main.tex
\documentclass{article}
\usepackage{spconf,amsmath,amssymb,graphicx,booktabs,array}
\usepackage[table]{xcolor}
\usepackage{tikz,hyperref,microtype,balance,iftex}
\ifPDFTeX\else
  \usepackage{fontspec}
\fi
\usetikzlibrary{arrows.meta,positioning,calc}
\definecolor{audio}{HTML}{E16A54}
\definecolor{answer}{HTML}{7C444F}
\definecolor{ours}{HTML}{005F02}
\definecolor{ourslight}{HTML}{EAF3EA}
\definecolor{control}{HTML}{C0B87A}
\definecolor{ink}{HTML}{172033}
\hypersetup{hidelinks}
\newcommand{\method}{PW-Pruning}
\newcommand{\inst}{PW-DO-ACP}
\newcommand{\best}[1]{\textbf{#1}}
\newcommand{\pw}{\textcolor{ours}{\bfseries PW}}
\newcommand{\twox}{$2\!\times$}
\newcommand{\fourx}{$4\!\times$}
\newcommand{\E}[1]{\mbox{E#1}}
\renewcommand{\arraystretch}{1.08}
\makeatletter
\renewcommand\section{\@startsection{section}{1}{\z@}%
 {-2ex \@plus -.5ex \@minus -.2ex}{.85ex \@plus .2ex}{\normalfont}}
\renewcommand\subsection{\@startsection{subsection}{2}{\z@}%
 {-1.35ex \@plus -.3ex \@minus -.2ex}{.45ex \@plus .1ex}{\normalfont}}
\long\def\@makecaption#1#2{\vskip 3pt
 \setbox\@tempboxa\hbox{#1. #2}%
 \ifdim\wd\@tempboxa>\hsize #1. #2\par\else
 \hbox to\hsize{\hfil\box\@tempboxa\hfil}\fi}
\makeatother
\title{PW-PRUNING: PHASE-AWARE EXPERT SELECTION\\
FOR SPEECH-CENTRIC OMNI-MODAL MODELS}
\name{\shortstack[c]{
Junjie Cao$^{1}$, Fengping Tian$^{1}$, Wenjie Zhou$^{1}$\\
Xuanfan Ni$^{1}$, Chenyang Lyu$^{*1}$, Siyou Liu$^{2}$, Longyue Wang$^{1}$, Weihua Luo$^{1}$}
\thanks{\fontsize{9}{11.25}\selectfont $^{*}$Corresponding author: Chenyang Lyu.}}
\address{$^{1}$Alibaba, Hangzhou, China \qquad $^{2}$University of Macau}
\begin{document}
\ninept
\setlength{\textfloatsep}{7pt plus 2pt minus 2pt}
\setlength{\dbltextfloatsep}{7pt plus 2pt minus 2pt}
\setlength{\floatsep}{6pt plus 2pt minus 2pt}
\setlength{\dblfloatsep}{6pt plus 2pt minus 2pt}
\setlength{\intextsep}{6pt plus 2pt minus 2pt}
\renewcommand{\dbltopfraction}{.85}
\maketitle

\begin{abstract}
Sparse routing reduces per-token computation but leaves the full expert bank
resident in memory. We study expert pruning for speech-centric omni-modal
models, where pooling calibration statistics can obscure the different expert
demands of acoustic-input processing and text-answer generation. We propose
\method{}, a phase-aware extension of DO-ACP that constructs separate
importance-weighted kernels for acoustic-prefill and teacher-forced answer
tokens, then combines phase-weighted log-determinant marginal gains to select one
static expert bank. On Qwen3-Omni, retaining 32 of 128 experts per layer
improves MMAU-mini and MMSU accuracy over matched pooled selection by
$3.20$ and $4.38$ percentage points, respectively, after supervised fine-tuning,
averaged over three runs. Single-phase and shuffled-label controls on this
public backbone suggest that semantic phase information contributes to these
gains. Experiments on a non-public Qwen3.5-Omni backbone and English-only
zero-shot evaluation of Nemotron provide complementary evidence.
Across the tested Qwen budgets, AU gains are larger under stronger pruning,
without inference-time mask switching.
\end{abstract}

\section{Introduction}

Expert pruning reduces the resident memory of mixture-of-experts (MoE)
models~\cite{kim2026dense,lasby2025reap}: sparse routing lowers per-token
computation but retains the full expert bank~\cite{switchtransformer,mixtral}.
Multimodal applications span speech/music generation~\cite{yang2026heartmula,han2025quantize,cao2025comprehend},
human animation~\cite{li2025dispose}, and video protection~\cite{cao2025videoguard,li2024uvcg}.
Omni-modal deployments such as speech assistants and translation may
prioritize speech-centric capabilities~\cite{qwen3omni,nemotron3omni},
motivating a smaller shared expert bank that preserves them.

Existing approaches use task-specific selection~\cite{chen2022taskspecific},
output reconstruction~\cite{lu2024experts}, or expert importance: PS ranks
routing mass~\cite{kim2026dense}, while REAP incorporates router-weighted
output norms~\cite{lasby2025reap}. DO-ACP additionally accounts for redundancy
among selected experts through activation-weighted log-determinant selection
~\cite{kim2026dense}. Yet pooling calibration statistics can obscure
\emph{where expert value arises}, even within one task. We distinguish two
token regions, or \emph{phases}, within a speech request: listening comprises
acoustic-prefill input positions that process audio evidence, while answering
comprises response positions that produce output text. Calibration uses
teacher-forced answers as a proxy for generation. An expert redundant in one
phase may complement the retained bank in the other; pooling before selection
can hide this distinction even with task-specific calibration.

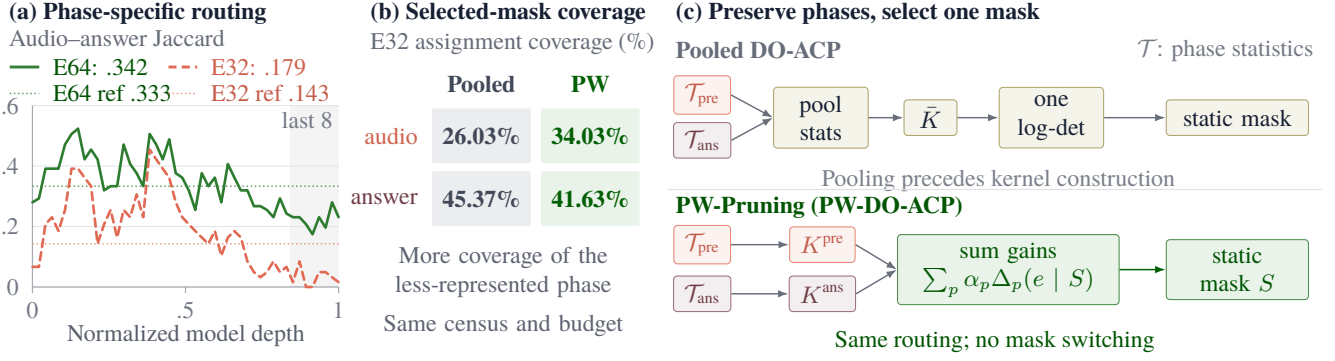
\begin{figure*}[t]
\centering
\input{figures/method_overview}
\caption{\textbf{From phase demand to one static mask.} Qwen3-Omni:
(a) audio-prefill/teacher-forced-answer frequency-mask Jaccard. Legends: means;
shading: last eight layers; dotted: $B/(2E-B)$ cardinality guides.
(b) Actual \E{32} mask assignment coverage, task/layer macro-averaged on calibration.
(c) Phase-wise selection. E: experts per MoE layer; normalized depth.}
\label{fig:method}
\end{figure*}

On Qwen3-Omni, the phases' high-frequency expert sets have mean Jaccard
overlap $.179$ at \E{32} and $.342$ at \E{64} (Fig.~\ref{fig:method}a).
Jaccard measures intersection over union: zero denotes disjoint sets and one
identical sets. Weak overlap also appears within tasks, motivating our
hypothesis that preserving phase identity during selection better represents
complementary demands than pooling statistics first. Our contributions are
threefold: (i)~we characterize this cross-phase routing disagreement;
(ii)~we propose \method{}, whose \inst{} selector constructs one
importance-weighted Gram kernel per phase and greedily combines
log-determinant marginal gains, retaining \emph{one static expert bank}
without runtime mask switching; and (iii)~we evaluate capability retention
through zero-shot pruning, matched recovery, and semantic controls on
Qwen3-Omni, with complementary evidence from Qwen3.5-Omni and Nemotron.

\section{Method}
\label{sec:method}

\subsection{Problem formulation}

Let $\theta$ be a pretrained omni model and
$S_\ell\subseteq[E_\ell]$ the experts retained in layer $\ell$.
The masks $\mathcal S=\{S_\ell\}_\ell$ define $\theta_{\mathcal S}$, inheriting
surviving weights and native routing fan-in. For task distributions $\mathcal D_g$ with nonnegative
weights $\pi_g$ summing to one, we express capability retention through the
reference loss $\mathcal L(\theta;x,y)=-\log p_\theta(y\mid x)/|y|$, with
reference length $|y|$:
\begin{equation}
\begin{aligned}
\underset{\mathcal S}{\operatorname{minimize}}\quad&
 \sum_g\pi_g\,\mathbb E_{\mathcal D_g}
 \bigl[\mathcal L(\theta_{\mathcal S};x,y)\bigr]\\
\text{subject to}\quad&
 S_\ell\subseteq[E_\ell],\quad |S_\ell|=B_\ell,\quad\forall\ell.
\end{aligned}
\label{eq:problem}
\end{equation}
Here $(x,y)$ is an input--reference pair. We fix $B_\ell=E_\ell/R$ above the
routing fan-in to compare selectors without changing layerwise capacity.
Cross-layer loss optimization is costly. We instead test the assumption that
retaining complementary phase-conditioned features improves capability retention,
using a layerwise log-det surrogate rather than a bound on Eq.~\ref{eq:problem}.
Below we suppress $\ell$, writing $E$ and $B$.

\subsection{From pooled importance to phase-wise coverage}

For token $t$, let $z_e(t)$ indicate whether the original router selects
expert $e$, $q_e(t)$ be its full-bank softmax probability before top-$k$
selection, and $\mathbf f_e(t)$ its ungated output. Acoustic-prefill and
teacher-forced-answer tokens define phases $p\in\{\mathrm{pre},\mathrm{ans}\}$.
The operator $\mathbb E_p$ averages tokens within each task group and then
weights nonempty groups equally. Applying DO-ACP~\cite{kim2026dense} within
each phase gives conditional routing confidence and an output Gram matrix:
\begin{equation}
\begin{aligned}
c_e^p &= \frac{\mathbb E_p[z_e q_e]}{\mathbb E_p[z_e]},\\
G_{ij}^p &= \mathbb E_p\bigl[\langle\mathbf f_i,\mathbf f_j\rangle\bigr].
\end{aligned}
\label{eq:stats}
\end{equation}
Set $c_e^p=0$ for an expert never selected in that phase. ACP importance
$u_e^p=c_e^p\sqrt{G_{ee}^p}$ combines conditional confidence with root-mean-square
output magnitude. With $D^p=\operatorname{Diag}(\sqrt{\mathbf u^p})$, the
importance-weighted kernel is
\begin{equation}
K^p=D^pG^pD^p\succeq0.
\label{eq:kernel}
\end{equation}
Thus $K^p_{ee}=c_e^p(G^p_{ee})^{3/2}$: the amplitude exponent is inherited from
DO-ACP, not derived from optimal design. Our contribution separates phases
while retaining this weighting. For nonnegative phase weights
$\alpha_p$ summing to one and regularization $\lambda_p>0$, define
\begin{equation}
\begin{aligned}
F(S)&=\sum_p\alpha_p\log\det\left(I+\frac{K^p[S,S]}{\lambda_p}\right),\\
S^\star&\in\underset{S\subseteq[E],\,|S|=B}{\arg\max}\;F(S).
\end{aligned}
\label{eq:objective}
\end{equation}
Here $K^p[S,S]$ is a principal submatrix and $I$ the matching identity.
This regularized feature-volume objective connects DPP diversity with
information-based sensor selection~\cite{kulesza2012dpp,krause2008sensor}.
For an auxiliary Gaussian covariance $K^p$ and noise $\lambda_p I$, each
utility equals twice the information gain~\cite{srinivas2010gp}, not the
language model's Fisher information. We use $\alpha_p=1/2$ to
balance phases, not to reproduce their token frequencies. Setting
$\lambda_p=\operatorname{tr}(K^p)/(BE)$ makes each utility invariant to
positive global kernel rescaling, while preserving relative expert energies.
Zero kernels use a positive floor. These are design choices, not optimality claims.

Pooled DO-ACP instead forms
\begin{equation}
\bar c_e=\frac{\sum_p\alpha_p\mathbb E_p[z_e q_e]}
 {\sum_p\alpha_p\mathbb E_p[z_e]},\qquad
\bar G=\sum_p\alpha_p G^p,
\label{eq:pooled}
\end{equation}
then constructs one kernel and ridge by the same rule. Conditional ratios,
kernel construction, and log-determinants do not commute with averaging, so generally
$F_{\rm pool}(S)\neq F(S)$ even with matched weights. PW changes
\emph{when} phase information is discarded, not the budget or routing.

\subsection{Greedy selection and deployment}

Starting from $S=\varnothing$, maintain each phase's Cholesky factor
$L_pL_p^\top=K^p_{SS}+\lambda_p I$, where subscripts denote matrix blocks.
A candidate $e\notin S$ contributes the Schur residual
\begin{equation}
v_p(e\mid S)=K^p_{ee}-\bigl\|L_p^{-1}K^p_{Se}\bigr\|_2^2,
\label{eq:residual}
\end{equation}
with the second term zero for empty $S$. It measures what the current subset
does not explain in phase $p$. The block determinant identity gives
\begin{equation}
\Delta(e\mid S)=\sum_p\alpha_p\log\left(1+\frac{v_p(e\mid S)}{\lambda_p}\right).
\label{eq:gain}
\end{equation}
We choose the largest $\Delta$ until $|S|=B$, evaluating $L_p^{-1}K^p_{Se}$
by triangular solves rather than explicit inversion. Monotone submodularity
gives greedy a $1-1/e$ approximation for $F$~\cite{nemhauser1978},
not a guarantee on language loss or benchmark accuracy.
The final subset serves both phases; subsequent recovery keeps its membership
fixed. Inference needs neither phase labels nor mask switching.

\section{Experiment}

In this section, we evaluate \method{} through zero-shot pruning, matched
recovery, semantic ablations, and cross-model transfer. These experiments test
our central hypothesis: retaining listening and answering as distinct selection
objectives improves capability preservation within a fixed expert budget.

\input{tables/main_results}

\begin{figure*}[t]
\centering\input{figures/au_recovery}
\caption{\textbf{Phase-semantics recovery on Qwen3-Omni, \E{32}.}
All methods use the same expert budget and SFT schedule. Final Pooled/PW
error bars show three-run sample s.d. (Table~\ref{tab:main}); mean control
endpoints are listed in Table~\ref{tab:ablation}.}
\label{fig:recovery}
\end{figure*}
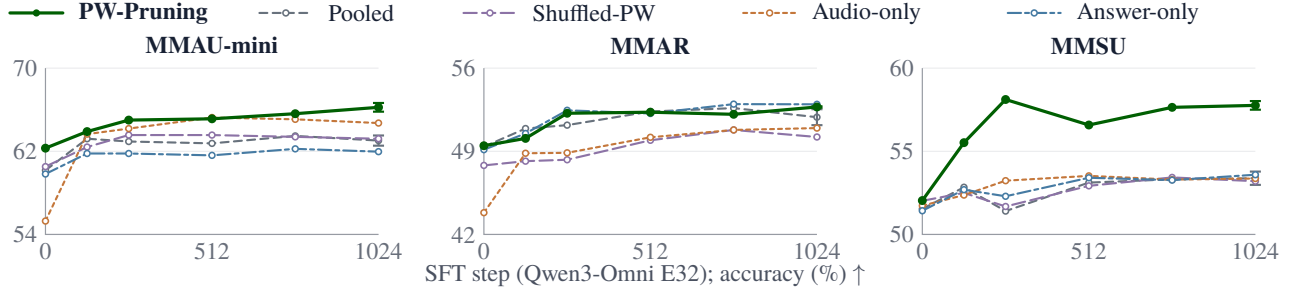

\input{tables/phase_ablation}
\input{tables/generalization}

\subsection{Experimental protocol}

\textbf{Models and census.} We evaluate text responses from
Qwen3-Omni-30B-A3B-Instruct~\cite{qwen3omni} (Thinker: approximately
30B total/3B active parameters, 48 layers, 128 experts per layer, top-8
routing~\cite{qwen3omni_config}). Pruning to
\E{64}/\E{32} removes expert tensors and router rows, preserving top-8 routing,
perception, and attention. Transfer uses Nemotron
3 Nano Omni-30B-A3B-Reasoning~\cite{nemotron3omni} at \E{64}/\E{32}, and a
non-public Qwen3.5-Omni backbone at $2\times/4\times$ expert-bank reduction.
Both Qwen censuses share 2,048 training-source requests balanced across ASR,
speech-to-text translation (S2TT), and audio understanding (AU). We count
acoustic-prefill and teacher-forced answers separately, excluding
instruction/special tokens. Stratified $1/64$ sampling estimates ungated Grams
for all experts. Nemotron uses English calibration and public English-ASR/AU only.

\textbf{Baselines.} We compare with \emph{Pooled DO-ACP}, which builds one
kernel from pooled statistics;
\emph{PS}~\cite{kim2026dense} ranks routed probability mass; \emph{REAP}
~\cite{lasby2025reap} ranks conditional router-weighted output norms from
its all-token census. Pooled shares PW's calibration, task/phase weights,
budget, regularization rule, and greedy solver. 

\textbf{Ablation.} Audio-only and Answer-only select from one phase's statistics
at the same budget. Shuffled-PW preserves two kernels and phase counts but
permutes labels within each request, testing the value of semantic grouping.
We apply these controls to both Qwen backbones under matched selection and
recovery settings; Table~\ref{tab:ablation} reports mean endpoint accuracies
over repeated recovery runs.

\textbf{Recovery.} All matched selectors use fixed sources of 6,638 ASR,
11,894 S2TT, and 16,758 AU examples with transcription, translation, and
audio-response supervision. Mixing targets equal expected supervised-answer-token
exposure per task. SFT uses 1,024 cross-entropy steps, global batch 32,
learning rate $2\cdot10^{-6}$, 5\% warmup, sequence cap 2,048, and no packing.
Retained language-model weights are tuned; perception encoders stay frozen.
KD uses reference-prefix KL at temperature 1, renormalized on
the teacher's top-64 vocabulary support, without CE mixing or student rollout.

\textbf{Public benchmarks.} Selectors share datasets and scoring within each
backbone. ASR comprises 54,494 examples from Common
Voice en/fr/yue/zh~\cite{ardila2020commonvoice}, LibriSpeech
clean/other~\cite{panayotov2015librispeech}, and OpenCpop~\cite{wang2022opencpop}. We retain native WER/CER families.
S2TT uses 1,104 FLEURS-derived examples in 234 directions~\cite{conneau2023fleurs}
with macro-direction SacreBLEU~\cite{post2018sacrebleu}.
AU uses 1,000 MMAU-mini, 1,000 MMAR, and 5,000 MMSU multiple-choice questions
~\cite{sakshi2024mmau,ma2025mmar,wang2026mmsu}. Greedy decoding uses
ASR/S2TT/AU caps of 512/256/128 tokens. Failures count as errors; metrics
are reported separately by task.

\subsection{Phase conditioning reallocates scarce capacity}

We first examine retained demand through routing preferences and selected
masks (Fig.~\ref{fig:method}). Qwen3-Omni phase-mask Jaccard averages $.179$
at \E{32} and $.342$ at \E{64}, falling to $.031$ and $.223$ in the final
eight layers. Within-task overlap is also low for ASR, S2TT, and AU
($J=.262$, $.232$, $.113$ at \E{32}). Task mixture alone therefore does not
explain the disagreement; Jaccard does not establish within-phase stability.
The selected masks show the allocation change: at the same \E{32} budget,
audio routing coverage rises from $26.03\%$ to $34.03\%$, while answer
coverage falls from $45.37\%$ to $41.63\%$ (Fig.~\ref{fig:method}b).
The gap narrows from $19.34$ to $7.60$ points, showing that PW reallocates
capacity toward underrepresented audio demand rather than raising both coverages.
Downstream tests assess whether this redistribution preserves capability.

\begin{figure}[!b]
\centering\input{figures/results_overview}
\caption{\textbf{Deployment savings.} H20, batch 1:
(a) peak memory relative to the unpruned model; (b) real-time factor.
Savings reflect expert budget, not PW-specific speedups.}
\label{fig:evidence}
\end{figure}
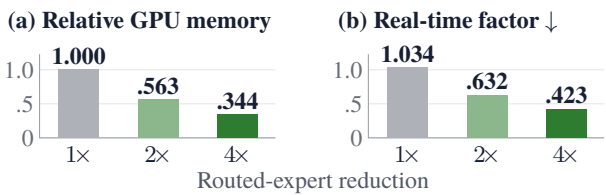

\subsection{Phase-aware selection improves capability retention}

We evaluate selection before and after adaptation through zero-shot,
SFT, and KD comparisons (Table~\ref{tab:main}). At \E{32}, PW lowers zero-shot
CV-WER from $27.0\%$ to $18.3\%$ and improves all three AU scores over
Pooled: the benefit is already present before recovery.

With matched SFT, PW improves six of eight means over Pooled. AU gains
are largest on MMAU-mini ($+3.20$ points) and MMSU ($+4.38$ points),
while MMAR changes more modestly. CV-C and OpenCpop CER remain higher
than Pooled ($14.25\%$ vs.\ $13.45\%$ and $5.05\%$ vs.\ $4.39\%$,
respectively); the OpenCpop gap also appears in the \E{32} zero-shot and
KD comparisons.
Crucially, Pooled already models importance and redundancy, using the same
expert budget and recovery resources. PW's advantage supports
retaining \emph{where an expert contributes},
rather than assessing its pooled value alone. PW also outperforms PS and
REAP on all eight SFT metrics, reducing CV-WER by $26.3\%$ and $24.5\%$,
respectively. The \E{32} KD comparison likewise favors PW on six metrics.
Thus, adapting surviving weights does not eliminate the importance of
phase-aware subset selection.

The AU gains become larger at tighter expert budgets. From \E{64} to
\E{32}, the SFT PW--Pooled gap grows from $2.08$ to $3.20$ points on
MMAU-mini and from $1.81$ to $4.38$ on MMSU. These larger gains support
the hypothesis that phase-specific functions become harder to replace as
spare capacity shrinks, making their preservation more valuable.

\subsection{Semantic phases preserve complementary capabilities}

To test phase semantics, Shuffled-PW preserves two kernels and phase counts
under matched budgets and recovery (Table~\ref{tab:ablation}). Qwen3-Omni PW
exceeds it on MMAU-mini and MMSU at all six reported checkpoints
(Fig.~\ref{fig:recovery}), gaining $3.02$ and $4.56$ points at step 1,024.
Qwen3.5-Omni gains $1.30$, $2.85$, and $.69$ points across the three
benchmarks, supporting semantic rather than arbitrary grouping.

Among single-phase controls, Audio-only leads MMAU-mini and Answer-only
MMAR on both backbones. Against the better single-phase control, Qwen3-Omni
PW gains $1.51$ and $4.17$ points on MMAU-mini and MMSU but loses $.22$
on MMAR. Qwen3.5-Omni PW gains $.25$ and $.21$ on MMAR and MMSU but
loses $.10$ on MMAU-mini. These complementary preferences support joint
phase preservation; the private-backbone differences remain small.

\subsection{Phase-aware selection transfers across backbones}

To assess transfer, we evaluate Nemotron before recovery and Qwen3.5-Omni
after SFT. On Nemotron, PW improves public English
ASR, MMAR, and MMSU at both budgets (Table~\ref{tab:nemotron}). At \E{32},
WER falls by $13.3\%$ and MMSU rises by $1.96$ points relative to Pooled,
with MMAU-mini unchanged. With no retraining,
these gains demonstrate transfer of the selection benefit independently of recovery.

On Qwen3.5-Omni, PW improves all eight reported means at $4\times$
expert-bank reduction (Table~\ref{tab:transfer}): CV-WER falls by $13.0\%$,
BLEU rises by $6.6\%$, and all three AU accuracies improve. Gains across
transcription, translation, and understanding after matched adaptation
complement Nemotron's zero-shot results, supporting the phase-allocation
principle across backbones and recovery settings.

\textbf{One static bank retains deployment savings.} Finally, H20 batch-1
profiles (64 ASR inputs, eight warmups, three repeats) show $65.6\%$ lower
peak memory and RTF $1.034\to.423$ at $4\times$ (Fig.~\ref{fig:evidence}).
These savings address the deployment motivation: PW retains one physically
pruned bank, without phase-dependent mask switching. The efficiency gain
comes from pruning; PW improves how the remaining capacity is allocated.

\section{Conclusion}

In this paper, we propose \method{} to prune omni-modal models for
speech-centric deployment. Phase-wise log-determinant gains preserve
listening and answering demands within one static mask. Zero-shot, recovery,
and semantic-control experiments support phase-aware capability retention,
especially at tighter budgets. Results on Qwen3-Omni, Qwen3.5-Omni, and
English-only Nemotron further demonstrate cross-model transfer.

\noindent\textbf{Limitations and future work.} Fixed budgets cannot exploit
depth-dependent phase overlap, while fixed weights and masks limit workload
adaptation. Future work could combine depth-aware allocation, workload-aware
weighting, and dynamic masks, subject to switching and memory costs.

\clearpage
\balance
\bibliographystyle{IEEEbib}
\bibliography{references}
\end{document}

%% file: figures/method_overview.tex
\begin{tikzpicture}[
  x=1cm,y=1cm,font=\fontsize{9}{11.25}\selectfont,
  every node/.style={text=ink},
  flow/.style={-{Latex[length=1.4mm,width=1.0mm]},line width=.55pt,draw=ink!65},
  axis/.style={draw=ink!38,line width=.45pt},
  grid/.style={draw=ink!12,line width=.35pt},
  box/.style={draw=ink!35,fill=white,rounded corners=1.4pt,
    minimum height=.47cm,inner xsep=4pt,inner ysep=3pt,align=center},
  pre/.style={box,draw=audio!80,fill=audio!9,text=audio!85!black},
  ans/.style={box,draw=answer!75,fill=answer!9,text=answer!90!black},
  pool/.style={box,draw=control!90!black,fill=control!16},
  pw/.style={box,draw=ours!70,fill=ourslight,text=ours!92!black}
]
\path[use as bounding box] (0,0) rectangle (17.6,4.75);

\node[anchor=west,font=\bfseries] at (0,4.56)
  {(a) Phase-specific routing};
\node[anchor=west,text=ink!66] at (0,4.23)
  {Audio--answer Jaccard};
\draw[draw=ours!80,line width=1.1pt] (.12,3.85) -- (.51,3.85);
\node[anchor=west,text=ours!90!black] at (.58,3.85) {E64: .342};
\draw[draw=audio,line width=1.1pt,densely dashed] (2.22,3.85) -- (2.60,3.85);
\node[anchor=west,text=audio!92!black] at (2.67,3.85) {E32: .179};
\draw[draw=ours!62,line width=.55pt,densely dotted] (.12,3.50) -- (.51,3.50);
\node[anchor=west,text=ours!85!black] at (.58,3.50) {E64 ref .333};
\draw[draw=audio!70,line width=.55pt,densely dotted] (2.22,3.50) -- (2.60,3.50);
\node[anchor=west,text=audio!88!black] at (2.67,3.50) {E32 ref .143};
\begin{scope}[shift={(.42,.94)},x={4.05cm/47},y=4cm]
  \fill[ink!5] (39.5,0) rectangle (47,.60);
  \foreach \yy/\lab in {0/0,.2/.2,.4/.4,.6/.6}{
    \draw[grid] (0,\yy) -- (47,\yy);
    \node[anchor=east,text=ink!60,inner sep=2pt] at (-1.175,\yy) {\lab};
  }
  \draw[axis] (0,.60) -- (0,0) -- (47,0);
  \foreach \xx/\lab in {0/0,23.5/.5,47/1}{
    \draw[axis] (\xx,0) -- (\xx,-.020);
    \node[anchor=north,text=ink!60,inner sep=3pt] at (\xx,-.021) {\lab};
  }
  \draw[draw=ours!62,line width=.55pt,densely dotted]
    (0,.3333333333) -- (47,.3333333333);
  \draw[draw=audio!70,line width=.55pt,densely dotted]
    (0,.1428571429) -- (47,.1428571429);
  \draw[draw=ours!82,line width=1.0pt,line join=round,line cap=round]
    plot coordinates {(0,0.28000000) (1,0.29292929) (2,0.39130435) (3,0.39130435)
    (4,0.39130435) (5,0.47126437) (6,0.50588235) (7,0.52380952) (8,0.42222222)
    (9,0.45454545) (10,0.42222222) (11,0.31958763) (12,0.33333333) (13,0.33333333)
    (14,0.47126437) (15,0.40659341) (16,0.37634409) (17,0.33333333) (18,0.50588235)
    (19,0.47126437) (20,0.42222222) (21,0.48837209) (22,0.37634409) (23,0.36170213)
    (24,0.31958763) (25,0.25490196) (26,0.37634409) (27,0.33333333) (28,0.36170213)
    (29,0.28000000) (30,0.40659341) (31,0.36170213) (32,0.31958763) (33,0.31958763)
    (34,0.26732673) (35,0.26732673) (36,0.25490196) (37,0.23076923) (38,0.29292929)
    (39,0.24271845) (40,0.23076923) (41,0.23076923) (42,0.20754717) (43,0.17431193)
    (44,0.23076923) (45,0.19626168) (46,0.28000000) (47,0.23076923)};
  \draw[draw=audio,line width=1.0pt,densely dashed,line join=round,line cap=round]
    plot coordinates {(0,0.06666667) (1,0.06666667) (2,0.20754717) (3,0.23076923)
    (4,0.18518519) (5,0.25490196) (6,0.39130435) (7,0.39130435) (8,0.36170213)
    (9,0.33333333) (10,0.14285714) (11,0.20754717) (12,0.25490196) (13,0.16363636)
    (14,0.25490196) (15,0.23076923) (16,0.30612245) (17,0.23076923) (18,0.45454545)
    (19,0.42222222) (20,0.39130435) (21,0.36170213) (22,0.28000000) (23,0.23076923)
    (24,0.20754717) (25,0.18518519) (26,0.16363636) (27,0.14285714) (28,0.18518519)
    (29,0.10344828) (30,0.16363636) (31,0.18518519) (32,0.16363636) (33,0.08474576)
    (34,0.04918033) (35,0.03225806) (36,0.04918033) (37,0.08474576) (38,0.04918033)
    (39,0.06666667) (40,0.01587302) (41,0.08474576) (42,0.00000000) (43,0.00000000)
    (44,0.04918033) (45,0.04918033) (46,0.03225806) (47,0.01587302)};
  \node[anchor=north east,text=ink!60,inner sep=2pt] at (47,.60)
    {last 8};
\end{scope}
\node[text=ink!72] at (2.45,.30) {Normalized model depth};

\node[anchor=west,font=\bfseries] at (4.78,4.56)
  {(b) Selected-mask coverage};
\node[anchor=west,text=ink!66] at (4.78,4.16)
  {E32 assignment coverage (\%)};
\node[text=ink!82,font=\bfseries] at (6.36,3.63) {Pooled};
\node[text=ours,font=\bfseries] at (7.80,3.63) {PW};
\node[anchor=east,text=audio!92!black] at (5.63,2.94) {audio};
\node[anchor=east,text=answer!90!black] at (5.63,2.11) {answer};
\fill[draw=white,fill=ink!8,rounded corners=1pt] (5.70,2.57) rectangle (7.02,3.31);
\fill[draw=white,fill=ourslight,rounded corners=1pt] (7.14,2.57) rectangle (8.46,3.31);
\fill[draw=white,fill=ink!8,rounded corners=1pt] (5.70,1.74) rectangle (7.02,2.48);
\fill[draw=white,fill=ourslight,rounded corners=1pt] (7.14,1.74) rectangle (8.46,2.48);
\node[text=ink!82,font=\bfseries] at (6.36,2.94) {26.03\%};
\node[text=ours,font=\bfseries] at (7.80,2.94) {34.03\%};
\node[text=ink!82,font=\bfseries] at (6.36,2.11) {45.37\%};
\node[text=ours,font=\bfseries] at (7.80,2.11) {41.63\%};
\node[align=center,text width=3.6cm,text=ink!70] at (6.65,1.14)
  {More coverage of the\\less-represented phase};
\node[align=center,text width=3.6cm,text=ink!70] at (6.65,.43)
  {Same census and budget};

\node[anchor=west,font=\bfseries] at (8.82,4.56)
  {(c) Preserve phases, select one mask};
\node[anchor=west,text=ink!66,font=\bfseries] at (8.82,4.07)
  {Pooled DO-ACP};
\node[anchor=east,text=ink!60] at (17.48,4.07)
  {$\mathcal T$: phase statistics};
\node[pre,minimum width=.75cm] (pa) at (9.28,3.48) {$\mathcal T_{\rm pre}$};
\node[ans,minimum width=.75cm] (py) at (9.28,2.88) {$\mathcal T_{\rm ans}$};
\node[pool,text width=.97cm] (ps) at (10.85,3.18) {pool\\stats};
\node[pool,minimum width=.70cm] (pk) at (12.30,3.18) {$\bar K$};
\node[pool,text width=1.10cm] (po) at (13.90,3.18) {one\\log-det};
\node[pool,text width=1.60cm] (pm) at (16.36,3.18) {static mask};
\draw[flow] (pa.east) -- (ps.west);
\draw[flow] (py.east) -- (ps.west);
\draw[flow] (ps) -- (pk);
\draw[flow] (pk) -- (po);
\draw[flow] (po) -- (pm);
\node[text=ink!57,anchor=north] at (13.2,2.60)
  {Pooling precedes kernel construction};
\draw[draw=ink!13,line width=.45pt] (8.82,2.23) -- (17.48,2.23);
\node[anchor=west,text=ours!90!black,font=\bfseries] at (8.82,1.99)
  {PW-Pruning (PW-DO-ACP)};
\node[pre,minimum width=.75cm] (wa) at (9.28,1.50) {$\mathcal T_{\rm pre}$};
\node[ans,minimum width=.75cm] (wy) at (9.28,.85) {$\mathcal T_{\rm ans}$};
\node[pre,minimum width=.75cm] (ka) at (10.87,1.50) {$K^{\rm pre}$};
\node[ans,minimum width=.75cm] (ky) at (10.87,.85) {$K^{\rm ans}$};
\node[pw,text width=2.65cm] (wg) at (13.33,1.17)
  {sum gains\\ $\sum_p\alpha_p\Delta_p(e\mid S)$};
\node[pw,text width=1.60cm] (wm) at (16.36,1.17)
  {static mask $S$};
\draw[flow] (wa) -- (ka);
\draw[flow] (wy) -- (ky);
\draw[flow] (ka.east) -- (wg.west);
\draw[flow] (ky.east) -- (wg.west);
\draw[flow,draw=ours] (wg) -- (wm);
\node[text=ours!90!black] at (13.15,.22)
  {Same routing; no mask switching};
\end{tikzpicture}

%% file: tables/main_results.tex
\begin{table*}[t]
\centering
\fontsize{9}{11.25}\selectfont
\caption{\textbf{Speech-centric pruning on Qwen3-Omni.} CV-W/CV-C:
Common Voice en+fr micro-WER / yue+zh micro-CER; LS: LibriSpeech WER;
OCP: OpenCpop CER (\%); AU: accuracy (\%). E: retained experts.
SFT: mean$\pm$s.d.; bold: better Pooled/PW mean.}
\label{tab:main}
\setlength{\tabcolsep}{3.8pt}
\renewcommand{\arraystretch}{1.0}
\begin{tabular}{clrrrrrrrr}
\toprule
Budget & Selector & CV-W$\downarrow$ & CV-C$\downarrow$ & LS$\downarrow$ & OCP$\downarrow$
& BLEU$\uparrow$ & MMAU$\uparrow$ & MMAR$\uparrow$ & MMSU$\uparrow$\\
\midrule
E128 & Teacher &6.17&6.42&2.11&1.49&29.34&78.00&70.30&71.04\\
\midrule
\multicolumn{10}{l}{\textit{Zero-shot}}\\
E64 & Pooled & \best{10.6}&\best{9.0}&\best{3.1}&2.9&\best{18.4}&72.2&\best{64.6}&\best{67.8}\\
\rowcolor{ourslight}E64 & \pw & 11.3&9.9&3.2&\best{2.2}&\best{18.4}&\best{73.4}&63.9&67.5\\
E64 & PS & 23.3&6.9&3.0&2.7&17.7&71.6&64.3&66.0\\
E64 & REAP & 22.2&10.7&3.2&3.9&15.2&72.2&63.5&66.7\\
E32 & Pooled & 27.0&30.7&7.7&\best{5.8}&5.9&60.2&49.4&51.5\\
\rowcolor{ourslight}E32 & \pw & \best{18.3}&\best{26.1}&\best{6.3}&6.1&\best{6.6}&\best{62.3}&\best{49.5}&\best{52.0}\\
E32 & PS & 79.0&16.4&7.7&6.3&3.3&62.2&51.4&49.8\\
E32 & REAP & 42.2&28.1&6.7&6.8&5.2&62.2&50.8&51.0\\
\midrule
\multicolumn{10}{l}{\textit{Supervised fine-tuning (SFT)}}\\
E64 &Pooled &9.88$\pm$.16&8.89$\pm$.11&\best{2.92$\pm$.06}&\best{1.91$\pm$.12}&24.63$\pm$.15&72.27$\pm$.06&66.40$\pm$.44&67.92$\pm$.14\\
\rowcolor{ourslight}E64 &\pw &\best{9.81$\pm$.05}&\best{8.86$\pm$.15}&2.95$\pm$.02&2.05$\pm$.06&\best{24.88$\pm$.44}&\best{74.35$\pm$.12}&\best{67.20$\pm$.15}&\best{69.73$\pm$.12}\\
E32 &Pooled &14.80$\pm$.04&\best{13.45$\pm$.86}&5.21$\pm$.05&\best{4.39$\pm$.32}&11.12$\pm$.04&63.03$\pm$.49&51.87$\pm$.67&53.38$\pm$.40\\
\rowcolor{ourslight}E32 &\pw &\best{14.17$\pm$.36}&14.25$\pm$.73&\best{5.14$\pm$.01}&5.05$\pm$.10&\best{12.69$\pm$.09}&\best{66.23$\pm$.42}&\best{52.73$\pm$.06}&\best{57.76$\pm$.26}\\
E32 & PS &19.23$\pm$.12&16.34$\pm$.56&7.12$\pm$.43&6.17$\pm$.62&10.91$\pm$.25&63.30$\pm$.74&51.48$\pm$.65&54.15$\pm$.06\\
E32 & REAP &18.77$\pm$.33&17.65$\pm$.31&6.55$\pm$.39&6.48$\pm$.22&11.56$\pm$.34&62.79$\pm$.58&52.10$\pm$.73&53.11$\pm$.44\\
\midrule
\multicolumn{10}{l}{\textit{Knowledge distillation (KD)}}\\
E64 & Pooled &9.31&8.10&2.89&2.25&24.11&76.20&69.30&68.86\\
\rowcolor{ourslight}E64 & \pw &9.59&7.57&2.93&2.06&24.22&77.50&68.60&69.42\\
E32 & Pooled &15.08&12.84&4.74&4.50&13.93&60.90&52.50&55.70\\
\rowcolor{ourslight}E32 & \pw &14.28&12.42&4.35&5.12&14.44&62.40&52.50&56.26\\
\bottomrule
\end{tabular}
\end{table*}

%% file: figures/au_recovery.tex
\begingroup
\definecolor{auPW}{HTML}{005F02}
\definecolor{auPooled}{HTML}{68727D}
\definecolor{auShuffled}{HTML}{8C73A6}
\definecolor{auAudio}{HTML}{C4773B}
\definecolor{auAnswer}{HTML}{467BA3}
\begin{tikzpicture}[x=1cm,y=1cm,font=\fontsize{9}{11.25}\selectfont,every node/.style={text=ink},
 axis/.style={draw=ink!45,line width=.45pt},
 grid/.style={draw=ink!10,line width=.35pt},
 auPWline/.style={draw=auPW,line width=1.2pt,solid,line cap=round,line join=round},
 auPoolLine/.style={draw=auPooled,line width=.8pt,dash pattern=on 3pt off 2pt,line cap=round,line join=round},
 auShuffleLine/.style={draw=auShuffled,line width=.8pt,dash pattern=on 5pt off 2pt,line cap=round,line join=round},
 auAudioLine/.style={draw=auAudio,line width=.8pt,dash pattern=on 1.5pt off 1.5pt,line cap=round,line join=round},
 auAnswerLine/.style={draw=auAnswer,line width=.8pt,dash pattern=on 5pt off 1.5pt on 1.5pt off 1.5pt,line cap=round,line join=round}]
\path[use as bounding box] (0,0) rectangle (17.4,3.82);
\draw[auPWline] (0.30,3.58) -- (1.00,3.58);
\fill[auPW] (.65,3.58) circle[radius=1.35pt];
\node[anchor=west,font=\fontsize{9}{11.25}\selectfont\bfseries] at (1.10,3.58) {PW-Pruning};
\draw[auPoolLine] (3.60,3.58) -- (4.30,3.58);
\draw[draw=auPooled,fill=white,line width=.6pt] (3.95,3.58) circle[radius=1.1pt];
\node[anchor=west] at (4.40,3.58) {Pooled};
\draw[auShuffleLine] (6.30,3.58) -- (7.00,3.58);
\draw[draw=auShuffled,fill=white,line width=.6pt] (6.65,3.58) circle[radius=1.1pt];
\node[anchor=west] at (7.10,3.58) {Shuffled-PW};
\draw[auAudioLine] (10.00,3.58) -- (10.70,3.58);
\draw[draw=auAudio,fill=white,line width=.6pt] (10.35,3.58) circle[radius=1.1pt];
\node[anchor=west] at (10.80,3.58) {Audio-only};
\draw[auAnswerLine] (13.50,3.58) -- (14.20,3.58);
\draw[draw=auAnswer,fill=white,line width=.6pt] (13.85,3.58) circle[radius=1.1pt];
\node[anchor=west] at (14.30,3.58) {Answer-only};
\node[font=\fontsize{9}{11.25}\selectfont\bfseries] at (2.97,3.15) {MMAU-mini};
\begin{scope}[shift={(0.77,.65)},x=.004296875cm,y=0.137500000cm]
\draw[grid] (0,0) -- (1024,0);
\node[anchor=east,inner sep=3pt,text=ink!65] at (0,0) {54};
\draw[grid] (0,8) -- (1024,8);
\node[anchor=east,inner sep=3pt,text=ink!65] at (0,8) {62};
\draw[grid] (0,16) -- (1024,16);
\node[anchor=east,inner sep=3pt,text=ink!65] at (0,16) {70};
\draw[axis] (0,16) -- (0,0) -- (1024,0);
\node[anchor=north,inner sep=3pt,text=ink!65] at (0,0) {0};
\node[anchor=north,inner sep=3pt,text=ink!65] at (512,0) {512};
\node[anchor=north,inner sep=3pt,text=ink!65] at (1024,0) {1024};
\draw[auPoolLine] plot coordinates {(0,6.17009) (128,9.20874) (256,8.94063) (512,8.76188) (768,9.47686) (1024,9.03000)};
\draw[draw=auPooled,fill=white,line width=.6pt] (0,6.17009) circle[radius=1.1pt];
\draw[draw=auPooled,fill=white,line width=.6pt] (128,9.20874) circle[radius=1.1pt];
\draw[draw=auPooled,fill=white,line width=.6pt] (256,8.94063) circle[radius=1.1pt];
\draw[draw=auPooled,fill=white,line width=.6pt] (512,8.76188) circle[radius=1.1pt];
\draw[draw=auPooled,fill=white,line width=.6pt] (768,9.47686) circle[radius=1.1pt];
\draw[draw=auPooled,fill=white,line width=.6pt] (1024,9.03000) circle[radius=1.1pt];
\draw[auPooled,line width=.8pt] (1024,8.54000) -- (1024,9.52000);
\draw[auPooled,line width=.8pt] (1007,8.54000) -- (1041,8.54000);
\draw[auPooled,line width=.8pt] (1007,9.52000) -- (1041,9.52000);
\draw[auShuffleLine] plot coordinates {(0,6.52507) (128,8.40976) (256,9.56204) (512,9.56083) (768,9.38808) (1024,9.20755)};
\draw[draw=auShuffled,fill=white,line width=.6pt] (0,6.52507) circle[radius=1.1pt];
\draw[draw=auShuffled,fill=white,line width=.6pt] (128,8.40976) circle[radius=1.1pt];
\draw[draw=auShuffled,fill=white,line width=.6pt] (256,9.56204) circle[radius=1.1pt];
\draw[draw=auShuffled,fill=white,line width=.6pt] (512,9.56083) circle[radius=1.1pt];
\draw[draw=auShuffled,fill=white,line width=.6pt] (768,9.38808) circle[radius=1.1pt];
\draw[draw=auShuffled,fill=white,line width=.6pt] (1024,9.20755) circle[radius=1.1pt];
\draw[auAudioLine] plot coordinates {(0,1.28903) (128,9.65263) (256,10.18346) (512,11.24750) (768,11.07488) (1024,10.71672)};
\draw[draw=auAudio,fill=white,line width=.6pt] (0,1.28903) circle[radius=1.1pt];
\draw[draw=auAudio,fill=white,line width=.6pt] (128,9.65263) circle[radius=1.1pt];
\draw[draw=auAudio,fill=white,line width=.6pt] (256,10.18346) circle[radius=1.1pt];
\draw[draw=auAudio,fill=white,line width=.6pt] (512,11.24750) circle[radius=1.1pt];
\draw[draw=auAudio,fill=white,line width=.6pt] (768,11.07488) circle[radius=1.1pt];
\draw[draw=auAudio,fill=white,line width=.6pt] (1024,10.71672) circle[radius=1.1pt];
\draw[auAnswerLine] plot coordinates {(0,5.81510) (128,7.78832) (256,7.78657) (512,7.60785) (768,8.23396) (1024,7.96470)};
\draw[draw=auAnswer,fill=white,line width=.6pt] (0,5.81510) circle[radius=1.1pt];
\draw[draw=auAnswer,fill=white,line width=.6pt] (128,7.78832) circle[radius=1.1pt];
\draw[draw=auAnswer,fill=white,line width=.6pt] (256,7.78657) circle[radius=1.1pt];
\draw[draw=auAnswer,fill=white,line width=.6pt] (512,7.60785) circle[radius=1.1pt];
\draw[draw=auAnswer,fill=white,line width=.6pt] (768,8.23396) circle[radius=1.1pt];
\draw[draw=auAnswer,fill=white,line width=.6pt] (1024,7.96470) circle[radius=1.1pt];
\draw[auPWline] plot coordinates {(0,8.30000) (128,9.89656) (256,11.00187) (512,11.12469) (768,11.61594) (1024,12.23000)};
\draw[draw=auPW,fill=auPW,line width=.65pt] (0,8.30000) circle[radius=1.35pt];
\draw[draw=auPW,fill=auPW,line width=.65pt] (128,9.89656) circle[radius=1.35pt];
\draw[draw=auPW,fill=auPW,line width=.65pt] (256,11.00187) circle[radius=1.35pt];
\draw[draw=auPW,fill=auPW,line width=.65pt] (512,11.12469) circle[radius=1.35pt];
\draw[draw=auPW,fill=auPW,line width=.65pt] (768,11.61594) circle[radius=1.35pt];
\draw[draw=auPW,fill=auPW,line width=.65pt] (1024,12.23000) circle[radius=1.35pt];
\draw[auPW,line width=.8pt] (1024,11.81000) -- (1024,12.65000);
\draw[auPW,line width=.8pt] (1007,11.81000) -- (1041,11.81000);
\draw[auPW,line width=.8pt] (1007,12.65000) -- (1041,12.65000);
\end{scope}
\node[font=\fontsize{9}{11.25}\selectfont\bfseries] at (8.77,3.15) {MMAR};
\begin{scope}[shift={(6.57,.65)},x=.004296875cm,y=0.157142857cm]
\draw[grid] (0,0) -- (1024,0);
\node[anchor=east,inner sep=3pt,text=ink!65] at (0,0) {42};
\draw[grid] (0,7) -- (1024,7);
\node[anchor=east,inner sep=3pt,text=ink!65] at (0,7) {49};
\draw[grid] (0,14) -- (1024,14);
\node[anchor=east,inner sep=3pt,text=ink!65] at (0,14) {56};
\draw[axis] (0,14) -- (0,0) -- (1024,0);
\node[anchor=north,inner sep=3pt,text=ink!65] at (0,0) {0};
\node[anchor=north,inner sep=3pt,text=ink!65] at (512,0) {512};
\node[anchor=north,inner sep=3pt,text=ink!65] at (1024,0) {1024};
\draw[auPoolLine] plot coordinates {(0,7.38422) (128,8.91393) (256,9.20075) (512,10.34804) (768,10.63486) (1024,9.87000)};
\draw[draw=auPooled,fill=white,line width=.6pt] (0,7.38422) circle[radius=1.1pt];
\draw[draw=auPooled,fill=white,line width=.6pt] (128,8.91393) circle[radius=1.1pt];
\draw[draw=auPooled,fill=white,line width=.6pt] (256,9.20075) circle[radius=1.1pt];
\draw[draw=auPooled,fill=white,line width=.6pt] (512,10.34804) circle[radius=1.1pt];
\draw[draw=auPooled,fill=white,line width=.6pt] (768,10.63486) circle[radius=1.1pt];
\draw[draw=auPooled,fill=white,line width=.6pt] (1024,9.87000) circle[radius=1.1pt];
\draw[auPooled,line width=.8pt] (1024,9.20000) -- (1024,10.54000);
\draw[auPooled,line width=.8pt] (1007,9.20000) -- (1041,9.20000);
\draw[auPooled,line width=.8pt] (1007,10.54000) -- (1041,10.54000);
\draw[auShuffleLine] plot coordinates {(0,5.80989) (128,6.16857) (256,6.28689) (512,7.92683) (768,8.79681) (1024,8.20215)};
\draw[draw=auShuffled,fill=white,line width=.6pt] (0,5.80989) circle[radius=1.1pt];
\draw[draw=auShuffled,fill=white,line width=.6pt] (128,6.16857) circle[radius=1.1pt];
\draw[draw=auShuffled,fill=white,line width=.6pt] (256,6.28689) circle[radius=1.1pt];
\draw[draw=auShuffled,fill=white,line width=.6pt] (512,7.92683) circle[radius=1.1pt];
\draw[draw=auShuffled,fill=white,line width=.6pt] (768,8.79681) circle[radius=1.1pt];
\draw[draw=auShuffled,fill=white,line width=.6pt] (1024,8.20215) circle[radius=1.1pt];
\draw[auAudioLine] plot coordinates {(0,1.83264) (128,6.83411) (256,6.86966) (512,8.17730) (768,8.79681) (1024,8.95268)};
\draw[draw=auAudio,fill=white,line width=.6pt] (0,1.83264) circle[radius=1.1pt];
\draw[draw=auAudio,fill=white,line width=.6pt] (128,6.83411) circle[radius=1.1pt];
\draw[draw=auAudio,fill=white,line width=.6pt] (256,6.86966) circle[radius=1.1pt];
\draw[draw=auAudio,fill=white,line width=.6pt] (512,8.17730) circle[radius=1.1pt];
\draw[draw=auAudio,fill=white,line width=.6pt] (768,8.79681) circle[radius=1.1pt];
\draw[draw=auAudio,fill=white,line width=.6pt] (1024,8.95268) circle[radius=1.1pt];
\draw[auAnswerLine] plot coordinates {(0,7.13564) (128,8.49797) (256,10.44955) (512,10.18106) (768,10.96905) (1024,10.95410)};
\draw[draw=auAnswer,fill=white,line width=.6pt] (0,7.13564) circle[radius=1.1pt];
\draw[draw=auAnswer,fill=white,line width=.6pt] (128,8.49797) circle[radius=1.1pt];
\draw[draw=auAnswer,fill=white,line width=.6pt] (256,10.44955) circle[radius=1.1pt];
\draw[draw=auAnswer,fill=white,line width=.6pt] (512,10.18106) circle[radius=1.1pt];
\draw[draw=auAnswer,fill=white,line width=.6pt] (768,10.96905) circle[radius=1.1pt];
\draw[draw=auAnswer,fill=white,line width=.6pt] (1024,10.95410) circle[radius=1.1pt];
\draw[auPWline] plot coordinates {(0,7.46708) (128,8.08439) (256,10.20088) (512,10.28906) (768,10.11269) (1024,10.73000)};
\draw[draw=auPW,fill=auPW,line width=.65pt] (0,7.46708) circle[radius=1.35pt];
\draw[draw=auPW,fill=auPW,line width=.65pt] (128,8.08439) circle[radius=1.35pt];
\draw[draw=auPW,fill=auPW,line width=.65pt] (256,10.20088) circle[radius=1.35pt];
\draw[draw=auPW,fill=auPW,line width=.65pt] (512,10.28906) circle[radius=1.35pt];
\draw[draw=auPW,fill=auPW,line width=.65pt] (768,10.11269) circle[radius=1.35pt];
\draw[draw=auPW,fill=auPW,line width=.65pt] (1024,10.73000) circle[radius=1.35pt];
\draw[auPW,line width=.8pt] (1024,10.67000) -- (1024,10.79000);
\draw[auPW,line width=.8pt] (1007,10.67000) -- (1041,10.67000);
\draw[auPW,line width=.8pt] (1007,10.79000) -- (1041,10.79000);
\end{scope}
\node[font=\fontsize{9}{11.25}\selectfont\bfseries] at (14.57,3.15) {MMSU};
\begin{scope}[shift={(12.37,.65)},x=.004296875cm,y=0.220000000cm]
\draw[grid] (0,0) -- (1024,0);
\node[anchor=east,inner sep=3pt,text=ink!65] at (0,0) {50};
\draw[grid] (0,5) -- (1024,5);
\node[anchor=east,inner sep=3pt,text=ink!65] at (0,5) {55};
\draw[grid] (0,10) -- (1024,10);
\node[anchor=east,inner sep=3pt,text=ink!65] at (0,10) {60};
\draw[axis] (0,10) -- (0,0) -- (1024,0);
\node[anchor=north,inner sep=3pt,text=ink!65] at (0,0) {0};
\node[anchor=north,inner sep=3pt,text=ink!65] at (512,0) {512};
\node[anchor=north,inner sep=3pt,text=ink!65] at (1024,0) {1024};
\draw[auPoolLine] plot coordinates {(0,1.53652) (128,2.83683) (256,1.40485) (512,3.11665) (768,3.31416) (1024,3.38000)};
\draw[draw=auPooled,fill=white,line width=.6pt] (0,1.53652) circle[radius=1.1pt];
\draw[draw=auPooled,fill=white,line width=.6pt] (128,2.83683) circle[radius=1.1pt];
\draw[draw=auPooled,fill=white,line width=.6pt] (256,1.40485) circle[radius=1.1pt];
\draw[draw=auPooled,fill=white,line width=.6pt] (512,3.11665) circle[radius=1.1pt];
\draw[draw=auPooled,fill=white,line width=.6pt] (768,3.31416) circle[radius=1.1pt];
\draw[draw=auPooled,fill=white,line width=.6pt] (1024,3.38000) circle[radius=1.1pt];
\draw[auPooled,line width=.8pt] (1024,2.98000) -- (1024,3.78000);
\draw[auPooled,line width=.8pt] (1007,2.98000) -- (1041,2.98000);
\draw[auPooled,line width=.8pt] (1007,3.78000) -- (1041,3.78000);
\draw[auShuffleLine] plot coordinates {(0,2.02165) (128,2.49710) (256,1.67974) (512,2.92249) (768,3.42742) (1024,3.20201)};
\draw[draw=auShuffled,fill=white,line width=.6pt] (0,2.02165) circle[radius=1.1pt];
\draw[draw=auShuffled,fill=white,line width=.6pt] (128,2.49710) circle[radius=1.1pt];
\draw[draw=auShuffled,fill=white,line width=.6pt] (256,1.67974) circle[radius=1.1pt];
\draw[draw=auShuffled,fill=white,line width=.6pt] (512,2.92249) circle[radius=1.1pt];
\draw[draw=auShuffled,fill=white,line width=.6pt] (768,3.42742) circle[radius=1.1pt];
\draw[draw=auShuffled,fill=white,line width=.6pt] (1024,3.20201) circle[radius=1.1pt];
\draw[auAudioLine] plot coordinates {(0,1.73057) (128,2.36768) (256,3.23207) (512,3.52113) (768,3.29798) (1024,3.36382)};
\draw[draw=auAudio,fill=white,line width=.6pt] (0,1.73057) circle[radius=1.1pt];
\draw[draw=auAudio,fill=white,line width=.6pt] (128,2.36768) circle[radius=1.1pt];
\draw[draw=auAudio,fill=white,line width=.6pt] (256,3.23207) circle[radius=1.1pt];
\draw[draw=auAudio,fill=white,line width=.6pt] (512,3.52113) circle[radius=1.1pt];
\draw[draw=auAudio,fill=white,line width=.6pt] (768,3.29798) circle[radius=1.1pt];
\draw[draw=auAudio,fill=white,line width=.6pt] (1024,3.36382) circle[radius=1.1pt];
\draw[auAnswerLine] plot coordinates {(0,1.42333) (128,2.69123) (256,2.29420) (512,3.40787) (768,3.26562) (1024,3.59035)};
\draw[draw=auAnswer,fill=white,line width=.6pt] (0,1.42333) circle[radius=1.1pt];
\draw[draw=auAnswer,fill=white,line width=.6pt] (128,2.69123) circle[radius=1.1pt];
\draw[draw=auAnswer,fill=white,line width=.6pt] (256,2.29420) circle[radius=1.1pt];
\draw[draw=auAnswer,fill=white,line width=.6pt] (512,3.40787) circle[radius=1.1pt];
\draw[draw=auAnswer,fill=white,line width=.6pt] (768,3.26562) circle[radius=1.1pt];
\draw[draw=auAnswer,fill=white,line width=.6pt] (1024,3.59035) circle[radius=1.1pt];
\draw[auPWline] plot coordinates {(0,2.03782) (128,5.51832) (256,8.11395) (512,6.58017) (768,7.64202) (1024,7.76000)};
\draw[draw=auPW,fill=auPW,line width=.65pt] (0,2.03782) circle[radius=1.35pt];
\draw[draw=auPW,fill=auPW,line width=.65pt] (128,5.51832) circle[radius=1.35pt];
\draw[draw=auPW,fill=auPW,line width=.65pt] (256,8.11395) circle[radius=1.35pt];
\draw[draw=auPW,fill=auPW,line width=.65pt] (512,6.58017) circle[radius=1.35pt];
\draw[draw=auPW,fill=auPW,line width=.65pt] (768,7.64202) circle[radius=1.35pt];
\draw[draw=auPW,fill=auPW,line width=.65pt] (1024,7.76000) circle[radius=1.35pt];
\draw[auPW,line width=.8pt] (1024,7.50000) -- (1024,8.02000);
\draw[auPW,line width=.8pt] (1007,7.50000) -- (1041,7.50000);
\draw[auPW,line width=.8pt] (1007,8.02000) -- (1041,8.02000);
\end{scope}
\node[text=ink!78,font=\fontsize{9}{11.25}\selectfont] at (8.7,.11)
 {SFT step (Qwen3-Omni E32); accuracy (\%) $\uparrow$};
\end{tikzpicture}
\endgroup

%% file: tables/phase_ablation.tex
\begin{table}[!t]
\centering
\fontsize{9}{11.25}\selectfont
\caption{\textbf{Phase-semantics ablation.} Mean AU accuracy (\%)
across recovery runs at SFT step 1,024. Qwen3-Omni \E{32} Pooled/PW:
Table~\ref{tab:main}; Qwen3.5-Omni $4\times$ Pooled/PW:
same means as Table~\ref{tab:transfer}.
Bold: highest mean per column in the Qwen3.5-Omni block.}
\label{tab:ablation}
\setlength{\tabcolsep}{4pt}
\renewcommand{\arraystretch}{1.0}
\begin{tabular}{@{}lrrr@{}}\toprule
Selector & MMAU$\uparrow$ & MMAR$\uparrow$ & MMSU$\uparrow$\\
\midrule
\multicolumn{4}{l}{\textit{Qwen3-Omni, \E{32} (Fig.~\ref{fig:recovery})}}\\
Shuffled-PW & 63.21&50.20&53.20\\
Audio-only & 64.72&50.95&53.36\\
Answer-only & 61.96&52.95&53.59\\
\midrule
\multicolumn{4}{l}{\textit{Qwen3.5-Omni, $4\times$ reduction}}\\
Pooled & 70.40&62.60&66.02\\
Audio-only & \best{72.60}&61.10&65.96\\
Answer-only & 69.80&62.80&66.24\\
Shuffled-PW & 71.20&60.20&65.76\\
\rowcolor{ourslight}\pw & 72.50&\best{63.05}&\best{66.45}\\
\bottomrule\end{tabular}
\end{table}

%% file: tables/generalization.tex
\begin{table*}[!t]
\centering
\fontsize{9}{11.25}\selectfont
\caption{\textbf{Transfer to Qwen3.5-Omni.} A non-public research backbone;
budgets are relative to the unpruned model. Matched SFT recovery:
mean$\pm$sample s.d. across repeated runs.
Metrics follow Table~\ref{tab:main}. Bold: better mean within each Pooled/PW pair.}
\label{tab:transfer}
\setlength{\tabcolsep}{3.5pt}
\renewcommand{\arraystretch}{1.0}
\begin{tabular}{clrrrrrrrr}
\toprule
$R$ &Selector &CV-W$\downarrow$&CV-C$\downarrow$&LS$\downarrow$&OCP$\downarrow$
&BLEU$\uparrow$&MMAU$\uparrow$&MMAR$\uparrow$&MMSU$\uparrow$\\
\midrule
$1\times$ &Teacher &5.68&5.70&2.26&1.06&30.72&78.50&73.60&71.10\\
\twox & Pooled &\best{9.67$\pm$.04}&8.77$\pm$.28&\best{3.01$\pm$.00}&\best{2.10$\pm$.03}&\best{26.14$\pm$.25}&75.70$\pm$.36&\best{70.40$\pm$.26}&\best{69.83$\pm$.21}\\
\rowcolor{ourslight}\twox & \pw &9.96$\pm$.15&\best{8.76$\pm$.48}&3.05$\pm$.03&2.64$\pm$.57&25.89$\pm$.03&\best{76.20$\pm$.62}&70.17$\pm$.71&69.49$\pm$.44\\
\fourx & Pooled &16.48$\pm$.22&14.75$\pm$.20&5.04$\pm$.11&5.22$\pm$.77&12.81$\pm$.23&70.40$\pm$.14&62.60$\pm$.57&66.02$\pm$.42\\
\rowcolor{ourslight}\fourx & \pw &\best{14.34$\pm$.48}&\best{14.73$\pm$1.08}&\best{4.66$\pm$.08}&\best{4.52$\pm$.41}&\best{13.65$\pm$.24}&\best{72.50$\pm$.71}&\best{63.05$\pm$1.06}&\best{66.45$\pm$.47}\\
\bottomrule
\end{tabular}
\end{table*}

\begin{table}[!t]
\centering
\fontsize{9}{11.25}\selectfont
\caption{\textbf{Zero-shot Nemotron transfer.}
English-subset public ASR WER and AU accuracy (\%);
E: retained experts. No recovery; bold: better Pooled/PW score.}
\label{tab:nemotron}
\setlength{\tabcolsep}{2pt}
\renewcommand{\arraystretch}{1.0}
\begin{tabular}{@{}clrrrr@{}}
\toprule
Budget & Selector & En-WER$\downarrow$ & MMAU$\uparrow$ & MMAR$\uparrow$ & MMSU$\uparrow$\\
\midrule
E128 & Teacher &3.378&77.60&60.70&62.60\\
\midrule
E64 & Pooled &4.173&\best{75.00}&58.50&60.00\\
\rowcolor{ourslight}E64 & \pw &\best{4.002}&74.30&\best{59.10}&\best{61.08}\\
E32 & Pooled &5.535&\best{69.50}&52.10&50.58\\
\rowcolor{ourslight}E32 & \pw &\best{4.797}&\best{69.50}&\best{52.50}&\best{52.54}\\
\bottomrule
\end{tabular}
\end{table}

%% file: figures/results_overview.tex
\begin{tikzpicture}[
  x=1cm,y=1cm,font=\fontsize{9}{11.25}\selectfont,
  every node/.style={text=ink},
  axis/.style={draw=ink!43,line width=.4pt},
  grid/.style={draw=ink!12,line width=.3pt},
  val/.style={font=\fontsize{9}{11.25}\selectfont\bfseries,inner sep=1.2pt}
]
\path[use as bounding box] (0,0) rectangle (8.4,2.55);

\node[anchor=west,font=\fontsize{9}{11.25}\selectfont\bfseries] at (.05,2.28)
  {(a) Relative GPU memory};
\node[anchor=west,font=\fontsize{9}{11.25}\selectfont\bfseries] at (4.42,2.28)
  {(b) Real-time factor $\downarrow$};
\foreach \offset in {.58,4.93}{
  \begin{scope}[shift={(\offset,.75)},x=1cm,y=.90cm]
    \foreach \yy/\lab in {0/0,.5/.5,1/1.0}{
      \draw[grid] (0,\yy) -- (3.12,\yy);
      \node[anchor=east,text=ink!65,inner sep=2pt] at (0,\yy) {\lab};
    }
    \draw[axis] (0,1.23) -- (0,0) -- (3.12,0);
    \foreach \xx/\lab in {.52/1,1.57/2,2.62/4}{
      \node[anchor=north,inner sep=3pt] at (\xx,0) {$\lab\!\times$};
    }
  \end{scope}
}
\begin{scope}[shift={(.58,.75)},x=1cm,y=.90cm]
  \fill[ink!35] (.25,0) rectangle (.79,1.000000);
  \fill[ours!45] (1.30,0) rectangle (1.84,.5628423436);
  \fill[ours!82] (2.35,0) rectangle (2.89,.3442635154);
  \foreach \xx/\yy/\lab in {.52/1/1.000,1.57/.5628423436/.563,2.62/.3442635154/.344}{
    \node[anchor=south,val] at (\xx,{\yy+.025}) {\lab};
  }
\end{scope}
\begin{scope}[shift={(4.93,.75)},x=1cm,y=.90cm]
  \fill[ink!35] (.25,0) rectangle (.79,1.0340642586);
  \fill[ours!45] (1.30,0) rectangle (1.84,.6319755515);
  \fill[ours!82] (2.35,0) rectangle (2.89,.4229204104);
  \foreach \xx/\yy/\lab in {.52/1.0340642586/1.034,1.57/.6319755515/.632,2.62/.4229204104/.423}{
    \node[anchor=south,val] at (\xx,{\yy+.025}) {\lab};
  }
\end{scope}
\node[text=ink!75] at (4.20,.16) {Routed-expert reduction};
\end{tikzpicture}